\documentclass[12pt,preprint]{aastex}

\received{2007 March 2}
\begin{document}
\pagenumbering{arabic}

\title{THE LUMINOSITY DISTRIBUTION OF GLOBULAR CLUSTERS IN DWARF GALAXIES}

\author{Sidney van den Bergh}
\affil{Dominion Astrophysical Observatory, Herzberg Institute of Astrophysics, National Research Council of Canada, 5071 West Saanich Road, Victoria, BC, V9E 2E7, Canada}
\email{sidney.vandenbergh@nrc.gc.ca}

\begin{abstract}
 The majority of the globular clusters associated with the
Sagittarius dwarf galaxy are faint. In this respect it differs significantly from  the globular cluster systems surrounding typical giant galaxies. The observation that most of globular clusters in the outer halo of the Galaxy are also sub-luminous may be understood by assuming that these clusters once also belonged to faint cluster-rich dwarf systems that were subsequently captured and destroyed by the Milky Way System.
\end{abstract}

\keywords{galaxies: dwarf - galaxies : starclusters - globular clusters: general}

\section{INTRODUCTION}

 It has been known for many years (Harris 1991, Harris 2001) that
most globular cluster systems have a roughly Gaussian luminosity distributions that peaks slightly above, $M_{v}$ = -7. The fact that the five globular clusters associated with the Fornax dwarf spheroidal have  $M_{v}$ = -7.1 $\pm$ 0.5 (van den Bergh 2000, p.224) appeared to support the notion that the luminosity distribution of globular clusters might be independent of parent galaxy luminosity. However, this conclusion was recently challenged by van den Bergh (2006), who used data by Sharina et al. (2005), to show that the luminosity distribution of typical globular clusters embedded in dwarf galaxies having $M_{v} > -16$ seemed to differ dramatically from that of the globular clusters surrounding giant galaxies having $M_{v} < -16$. In particular it was found that the luminosity distribution of globular clusters in some dwarf galaxies appeared to increase monotonically down to the completeness limit of the data at $M_{v} \sim -5$.

\section{THE SAGITTARIUS DWARF}

Recently Jord\'{a}n et al. (2007) have pointed out that the results by
van den Bergh (2006) are mainly based on the data by Sharina et al. (2005), who may not have sufficiently taken into account the potential contamination of their list of candidate globular clusters.

Their results must therefore be regarded as being quite uncertain until cluster candidates have been confirmed spectroscopically. Fortunately such a check is already available using the globular clusters that appear to be associated with the nearby Sagittarius dwarf. Table 1 lists information on these clusters drawn from recent compilations by van den Bergh \& Mackey (2004) and Mackey \& van den Bergh (2005), supplemented by information on the newly discovered object Whiting 1 (Carraro et al. 2007). For the globulars listed in this table one finds $<M_{v}> = -5.8$  (or $<M_{v}> = -5.2$ if NGC 6715 = M54 (which may be the nucleus of the Sagittarius dwarf) is excluded. In other words the globulars associated with the Sagittarius dwarf are clearly less luminous than those surrounding typical giant galaxies. A Kolmogorov-Smirnov test shows that there is only a 3\% probability that the globular clusters in Sagittarius, and those in the main body of the Galaxy with Galactocentric distances $<$ 15 kpc (van den Bergh 2006), were drawn from the same parent distribution of luminosities. If M54 is excluded from the sample then an S-K test gives a probability of only 1\% that the Galactic and Sagittarius globular cluster samples were drawn from a similar parent distribution. This result suggests that the Sagittarius dwarf globular cluster resembles those typical of the cluster systems surrounding faint galaxies that were studied by van den Bergh (2006).

\section{DISCUSSION}

     It would be important to know if the luminosity
distribution of globular clusters is universal, or if it is a function of parent galaxy luminosity (mass). Analysis of the data presented by Sharina et al. strongly suggests that the luminosity distribution of globular clusters in galaxies with $M_{v} > -16$ differs dramatically from that of the much more thoroughly studied globular cluster systems surrounding massive galaxies with $M_{v} < -16$. That this difference is not an artifact produced by distance-dependent selection effects is shown by the fact that the Sagittarius system (which is the nearest known dwarf) exhibits the same type of globular cluster luminosity distribution that is observed in globular cluster systems associated with more distant dwarf galaxies. In this connection it is of interest to note that both dwarf galaxies (Tolstoy et al. 2003), and some of the globular clusters associated with them (Sbordone et al. 2005), have metallicity signatures that differ significantly from those encountered in more luminous systems. Both the dwarf spheroidals and the globulars associated with them are found to have the same low alpha-element to iron ratio and the same low Ni/Fe ratio. It is of interest to note (Sbordone 2005) that the outer halo cluster ($R_{gc}$ = 18.5 kpc) Ruprecht No. 106 shares the same peculiar metallicity signature.  This suggests that the anomalously low luminosity of many of the globular clusters in the outer Galactic halo (van den Bergh \& Mackey 2004) may be due to the fact that these objects were stripped from dwarf galaxies that have subsequently been tidally disrupted.
 
In summary it appears that both the anomalous luminosity
distribution of globular clusters in the outer halo, and their metallicity signature, might be understood by assuming that these objcets originated in dwarf galaxies that were subsequently captured and/or disrupted by Galactic tidal forces.

I am particularly indebted to the anonymous referee for
hinting at the possible evolutionary connection between the anomalous luminosity distribution of outer halo globulars and their anomalous abundance signature.

          
\begin{deluxetable}{lr}
\tablewidth{0pt}        
\tablecaption{Globular clusters that are probably associated with the Sagittarius dwarf galaxy} 

\tablehead{\colhead{Name} & \colhead{$M_{v}$}}

\startdata

Pal. 2      &      -8.01 \\
NGC 4147    &      -6.16  \\
NGC 6715    &     -10.01  \\
Ter. 7      &      -5.05  \\
Arp 2       &      -5.29  \\
Ter. 8      &      -5.05  \\
Pal. 12     &      -4.48  \\
Whi. 1      &     -2.42  \\

\enddata
\end{deluxetable}

\end{document}